\documentclass[%
 aip,
 jcp,%
 amsmath,amssymb,
floatfix,
 reprint,%
]{revtex4-1}

\usepackage{graphicx}
\usepackage{dcolumn}
\usepackage{bm}
\usepackage{capt-of}
\usepackage{pstricks}
\usepackage{amsfonts}
\usepackage[sort&compress]{natbib}

\usepackage{amsmath}
\usepackage{mathrsfs}
\DeclareMathAlphabet\mathbfcal{OMS}{cmsy}{b}{n}

\usepackage{subcaption}

\newcommand{\ch} {ECH~}
\newcommand{\chlong} {Energy Convex Hull}
\providecommand{\e}[1]{\ensuremath{\times 10^{#1}}}

\begin{document}


\title{Strongly bent double-stranded DNA: reconciling theory and experiment}

\author{Aleksander V. Drozdetski}
\thanks{These two authors contributed equally}
\affiliation{Department of Physics, Virginia Tech, Blacksburg, VA 24061, USA}
\author{Abhishek Mukhopadhyay}%
\thanks{These two authors contributed equally}
\affiliation{Department of Physics, Virginia Tech, Blacksburg, VA 24061, USA}
\author{Alexey V. Onufriev}
\email{alexey@cs.vt.edu}
\affiliation{Department of Physics, Virginia Tech, Blacksburg, VA 24061, USA}
\affiliation{Department of Computer Science, Virginia Tech, Blacksburg, VA 24061, USA}%
\affiliation{Center from Soft Matter and Biological Physics, Virginia Tech, Blacksburg, VA 24061, USA}%

\date{\today}

\begin{abstract}

The strong bending
of polymers is poorly understood. 
We propose a general quantitative framework of polymer bending that
includes both the weak and strong bending regimes on the same footing, 
based on a single general physical principle.
As the bending deformation increases beyond a certain (polymer-specific) point,
the change in the convexity properties of the effective bending energy of the
polymer makes the harmonic deformation energetically unfavorable: in this
strong bending regime the energy of the polymer varies linearly with the
average bending angle as the system follows the convex hull of the deformation
energy function. For double-stranded DNA, the effective bending deformation
energy 
becomes non-convex for  bends greater than
$\sim2^\circ$ per base-pair, equivalent to the curvature of a closed circular loop of
$\sim 160$ base pairs. A simple equation is derived
for the polymer loop energy that covers both the weak and strong bending
regimes. The theory shows quantitative agreement with  recent DNA cyclization
experiments on short DNA fragments, while maintaining the expected agreement
with experiment in the weak bending regime. Counter-intuitively, cyclization
probability ({\it j-factor}) 
of very short DNA loops is predicted to increase with
decreasing loop length;  the {\it j-factor} reaches its minimum
for loops of $\simeq45$ base pairs.   
Atomistic simulations reveal that
the attractive component of  the short-range
Lennard-Jones interaction between the backbone atoms can
explain the underlying
non-convexity of the DNA effective bending energy, leading to the linear bending
regime. Applicability of the theory to protein-DNA complexes, including the
nucleosome, is discussed.
\end{abstract}

\maketitle


\section{Introduction}

Deformation of polymers is ubiquitous, elastic properties of these
macromolecules are crucial for their dynamics. Biopolymers are abundant in
nature and play vital roles in many biological
processes~\cite{Grosberg,Garcia2007,Bustamante04,Nelson99}, which not only
depend upon the polymer structure, but also their physical
properties~\cite{Van2002,Gosline2002,Kasas2004}. Among biopolymers, DNA stands
out as a case of its own. Understanding DNA deformation is crucial for the
mechanistic grasp of vital cellular functions such as packaging of DNA
compactly into viral capsids, the chromatin, formation of protein/DNA complexes and regulation of gene expression~\cite{Garcia2007,Maher2010}. An all-important example of DNA
deformation, relevant to a variety of biological processes that depend on its
elastic properties, is DNA looping, which occurs in many
prokaryotic~\cite{Gralla1991} and eukariotic~\cite{Richmond2003} systems. A
number of regulatory proteins can loop DNA into various bent conformations,
critical for regulation of many biological processes involving
DNA~\cite{Schleif1992}. Most notably, DNA is strongly bent in the
nucleosome~\cite{Kornberg1974,Luger1997}, which is the fundamental unit of
genome packing: accessibility to genomic
information in eukareotes is modulated by the strength of DNA-protein
association~\cite{Henikoff07,Fenley2018}. Note that 
the majority of eukariotic genomic DNA (75-80\%) is packed
tightly into nucleosomes~\cite{Richmond2003}. 
Nanostructures made directly of DNA\cite{MaffeoAksimentiev2016} 
or those that use  DNA as a scaffold\cite{Mastroianni2009}, 
can be influenced\cite{ParkMirkin2008} 
by its mechanical properties on short length scales, providing yet
another impetus to understand the strong bending regime of the DNA.  

Experimental evidence on cyclization of DNA fragments shorter than $\sim$100 base-pairs
points to the fact that strongly bent DNA -- most relevant from a biological perspective
-- is considerably more flexible than expected from established models (worm-like chain)
that work well within the weak bending regime.
Yet, despite decades of  experimental and theoretical effort, the
story of how this arguably most important polymer behaves under deformation is
far from complete, with controversies and new developments
abound~\cite{Ha2012,Vologodskii2013,Savin2013,Zoli2018,Sivak2012,Salari2015,WuTan2015}. 
The current state and the
relevant terminology are briefly reviewed below.


	The bending flexibility of a polymer is conventionally quantified in terms
of its persistence length, $L_p$, a length scale below which the polymer
behaves more or less like a rigid rod. Specifically, $L_p$ is defined as length
of the polymer segment over which the time-averaged orientation of the polymer
becomes uncorrelated; for fragments smaller than $L_p$, the thermal fluctuation
alone is not enough to induce significant ($\sim$ 1 rad)
bending~\cite{Travers2004}. Here we use this definition of $L_p$ to
qualitatively separate the two bending regimes: if no significant bending
is observed on length scales shorter than $L_p$, the polymer can be deemed {\it
weakly} bent; otherwise the bending is assumed to be {\it strong}.

	For double-stranded DNA, a variety of experimental
techniques~\cite{Baumann1997,Crothers1992,Crothers1992,Vologodskii2002,Baldwin1981,Hagerman1988,Du2005},
revealed that $L_p \approx 150 $ bp or 500~\AA.
Based on the $L_p$ value and the above definition of strong bending, we
conclude that most of the DNA in eukareotes is strongly bent. Indeed, since the
nucleosome contains a stretch of double-stranded DNA of $\sim 150$ bp looped
almost twice, the DNA in this complex can be considered as strongly bent.

	Response of DNA to mechanical stress has been studied extensively
~\cite{Frank1985,Baldwin1981,Marko1994,Baumann1997,Bloomfield1997,Garcia2007,Lankas2000,Cloutier2004,
Wiggins2006,Wiggins2005,Biswas2012,Bomble2008,Cloutier2005,
Du2005,Lankas2006,Mazur2007,Seol2007,Strauss1994,Prevost2009,Maher2010,Fields2013,MathewHarbury2008},
leading to a consensus in modeling the weak bending regime.
Arguably the most widely used
simplified model of DNA bending is the worm-like chain (WLC) model. In the
original WLC~\cite{Porod1949,Du2005}, the polymer is modeled as a continuous,
isotropic elastic rod with its deformation energy being a quadratic function of
the deformation angle. In the discrete version of WLC model, the bending
energy of the polymer consisting of $N$ segments of 
length $l$ is given by:
\begin{equation}
E_{chain}  =\sum_i^{N-1}\frac{1}{2}k_B T \frac{L_p}{l} \theta_i^2
\label{eqn:ewlc}
\end{equation}
where $\theta_i$ is the angle between two consecutive segments (see inset of
Fig.~\ref{fig:conhull}).
While this simplistic model lacks some features of the real DNA, such as
sequence dependence of its local mechanical properties, it nevertheless
captures the key physics of weak polymer bending, which explains why the model
is robust and is widely adopted to interpret experiment. Various theoretical
models of DNA bending, including those that explicitly account for the
sequence-dependence~\cite{Coleman2003,Lankavs2003,Dixit2005,Fujii2007}, were
consequently developed that also assumed harmonic (quadratic) angular
deformation energy of DNA. There is very little doubt that the Hookean,
``elastic rod'' models accurately describe many polymers in the weak bending
regime~\cite{Grosberg}, including the double-stranded DNA\cite{Du2005,Mazur2007}.
Indeed, lowest order term of a Taylor series expansion of any well-behaved
function around its local minimum is quadratic, which means that for small
deviations from equilibrium, the response function can be considered harmonic.
However, by the same logic it should be expected that beyond a certain
threshold the bending energy may no longer be approximated by the quadratic
term alone;
investigations of possible influence of non-harmonic terms on the mechanical
properties of double-stranded DNA is a relatively new area. Historically, only
very large fragments (hundreds to thousands of base-pairs) were
investigated~\cite{Baumann1997,Crothers1992}, which are well described by the
WLC regardless of what happens on short length-scales~\cite{Wiggins2006}.


	However, within the last decade or so, the prevailing view of DNA as a
Hookean polymer was challenged by experiments that were able to investigate the
flexibility of DNA on scales smaller than several $L_p$.
	Counter-intuitively, small DNA fragments ($\approx100$bp) were found to have
much higher probability of cyclization (spontaneous formation of loops) than
that predicted by the WLC theory~\cite{Cloutier2005}. This discovery sparked
considerable controversy, which still remains unresolved. What is particularly
puzzling is that strongly bent DNA appears {\it less} rigid than the DNA in the
Hookean regime. Some of the follow-up experimental and theoretical work
supports the validity of WLC even for tightly bent
DNA~\cite{Du2005,Vologodskii2013,Mazur2014}, while others still show that
short, tightly bent  DNA is much more flexible~\cite{Wiggins2006,czapla2006,Ha2012}
than previously thought, in a manner that can not be described by a harmonic
model~\cite{Wiggins2006}. 

	Several theoretical models have been proposed to account for the unexpectedly
high flexibility of strongly bent double-stranded DNA. 
One popular model\cite{YanMarko2004} -- the meltable WLC or MWLC -- 
postulated that the extra flexibility
stems from formation of small local ``bubbles" of single stranded DNA, which is
much softer than the double helix.  However, the degree of softening provided
by the mechanism alone was later found\cite{FortiesPoirier2009} to be inadequate to
fully explain the very sharp bends in DNA observed experimentally; in atomistic
simulations, negative super-coiling was required to induce such bubbles in DNA
mini-circles\cite{Mitchell2011}.
	An early model\cite{CrickKlug1975}, put
forward well before the unusual DNA flexibility was discovered experimentally,
suggested that the energy of a bent double-helix could be lowered by formation
of sharp, $\sim 90^o$ kinks that maintain the Watson-Crick pairing along the
helix. Sharp kinks were indeed observed in a pioneering atomistic
simulation\cite{Lankas2006} some thirty years later,
but subsequent improvements in the simulation 
methodology indicated that these were only
induced at a high bend angle equivalent to those occurring in circles of just
$45$ base-pairs\cite{Curuksu2009}, while experimental softening of the DNA is
seen experimentally for circles as large as $\sim 106$ base-pairs\cite{Ha2012}.
Sharp kinks in
double-stranded DNA can be introduced empirically into the WLC model, 
{\it e.g.} by adding freely-bending hinge elements to the WLC chain, leading to
a kinkable WLC, or KWLC\cite{Wiggins2005}. A non-linear empirical bending
potential that allows for the possibility of $\sim 90^o$ kinks in double-stranded DNA
was recently proposed\cite{Vologodskii2013}, but its physical origins, the critical
value of the DNA curvature at which the kink occurs, and the corresponding
energy gain remained unknown\cite{Vologodskii2013}. At the same time, a purely
linear empirical bending potential was shown\cite{Wiggins2006} to describe the
softer DNA seen in AFM experiments, although the origin of the linear regime
and its parameters ({\it e.g.} critical bend angle where the linear regime
begins ) remained unclear. Are the kinking and the linear regime just two
manifestations of a deeper underlying principle?

	In summary, the nature of the effective bending energy of
double-stranded DNA in the strong bending regime as well as the precise
connection to the observed softening of the polymer is not fully clear. The
influence of mechanical constraints on this connection remains unexplored. It
is unclear how the softening of strongly bent DNA stems from
its atomic-level structure and interactions. From a more philosophical
standpoint, is hard to believe that
very special models are needed to describe the bending of the DNA; rather it
is more like that the curious case of the DNA is just
a special case of a broader underlying
theory applicable to all polymers.

	In this work we propose, and verify against available experiment, a unified
theoretical description of polymer bending that treats the weak and strong
bending regimes on the same footing, guided by a simple physical principle.
The proposed framework does not rely on {\it ad-hoc} postulates; instead, it
shows how the apparent softening of strongly bent DNA follows naturally from a
specific mathematical property of the experimentally-derived bending
energy. Simulations suggest an atomistic explanation for the specific
shape of the bending energy function.

\section{Methods}

\subsection{DNA bending energy from experimental data}
	A statistically significant, diverse set of several hundred PDB structures
of protein-DNA complexes was investigated previously in Ref. \cite{Du2005}.
The probability distribution of the experimental DNA bending angles was used in
Ref.~\cite{Du2005} to approximate the bending energy $E(\theta)$ (per base pair) as a
fourth oder polynomial: $E(\theta) = 203.1\theta^{2} - 552.7\theta^{3} + 416.8\theta^{4}$
(where $\theta$ is in radians and $E(\theta)$ is in units of $kT$). 
Here we use this $E(\theta)$ to represent the
experimental effective bending energy of the double-stranded DNA, blue line in
Fig.~\ref{fig:model_idea}.

\subsection{Atomistic MD simulations of closed DNA loops}

To avoid ``end effects", and make a close connection with DNA
cyclization experiments, we employed closed DNA circles to estimate their
effective bending energy $E(\theta)$ per base pair. 
DNA circles of various sizes (50-400 bp) were generated using
NAB\cite{Macke1998} (AmberTools) for sequence 
poly(dA).poly(dT),  helical repeat of 10 bp, and other 
parameters of B-DNA as specified in NAB.  
We deliberately chose this simple, uniform sequence to
focus on the basic physics of DNA deformation. 

All of the atomistic MD simulations were performed within AMBER-12 package, 
using ff99bsc0 force-field.     
The  Generalized Born (GB-HCT, AMBER option
igb=1) implicit solvation model was used to treat solvation effects, including 0.145M of
monovalent salt. No long-range cut-off was employed.  The model's 
performance in atomistic simulations of DNA, 
including studies of its deformation\cite{Bomble2008}, 
is well established\cite{Onufriev2010}. Two critical advantages of the implicit solvation 
over the more traditional explicit solvation\cite{OnufrievWaterReview2018} 
made the former the method of choice 
in this work. These are the superior  simulation efficiency for large DNA 
structures\cite{Anandakrishnan2015} and the 
straightforward manner in which their energies, including free energy of solvent re-arrangement, can be estimated\cite{Onufriev2010} within the implicit solvation framework.

	All DNA circles were
initially minimized for 1000 steps with ``P" atoms restrained
their original positions with a force
constant of 1.0 kcal/mol/\AA{}$^{2}$ to enforce the circular shape. Each system
was then heated to 300K and equilibrated for 100 ps with the same restraints as
for the minimization. Shake 
was used to constrain the hydrogen atoms; we employed 2 fs time-step for
the atomistic simulations.  
Finally, we generated 1 ns long MD trajectory for each
circle at 300K, with ``P" atoms also restrained with a force constant of 0.1
kcal/mol/\AA{}$^{2}$, sufficient to support the near perfect circular shape of
the fragment, but allowing for local re-arrangements. 
The energies and their components, including the
electrostatic, VDW, bond, etc.  were saved every 20 fs, 
and averaged over the whole
trajectory.  The relatively short simulation time allowed us to simulate even
the largest of the circles; it is justified by the use of the strong positional
restraints, which permit only local, very fast structural re-arrangements. For
smaller circles we verified that increasing the simulation time by an order of
magnitude had negligible effect on the computed averages. In Fig.
\ref{fig:md_energy}, the energy per bp 
was computed as the difference between per bp
potential energies of the given circle and the largest circle simulated, 
which is virtually unbent.

\subsection{Coarse-grained simulations of DNA loops} 

ESPResSo~\cite{Limbach06} was used to create and simulate  
coarse-grained closed loops of DNA
of different sizes, from  6 to 600 bp long. A single bead of the appropriate 
mass represents one base-pair of B-DNA; 
the bead-bead distance was set to $3.3$ \AA~, corresponding to the average distance between base pairs in canonical DNA.  
The bonds between the beads were made virtually 
inextensible
(very large coefficient of the quadratic bond stretching energy); the bond angle potential 
(effective bending energy) between neighboring beads was defined to have the same form  as in Fig.~\ref{fig:model_idea}, 
that is correspond to the bending potential inferred 
from the experimental data\cite{Du2005}. No further bead-bead interactions or
constraints on the loop geometry were imposed. The loops were 
simulated at $T=300K$, and energy-minimized using steepest descent.

\subsection{Coarse-grained simulations of confined DNA fragments}
\paragraph{Protein-DNA complex.}
ESPResSo~\cite{Limbach06} was used to create and simulate a 20
bead long fragment of ``DNA" bound to a spherical charged "protein",
Fig.~\ref{fig:confine}. The beads and their interactions were set up 
as described above, with the following modifications. 
The end beads were not linked to create a loop.  
Each bead carried a unit charge $q_s=$-1;
The bead charges
interacted only with a positive charge $Q$ of the ``protein", represented by a
spherical impenetrable constraint of radius $R$. In addition, two impenetrable
walls were placed above and below the charge $Q$ to minimize out-of-plane
bending of the ``DNA". The confining charge $Q$ was varied from 10 to 1000,
effectively sampling two orders of magnitude of confinement strength (defined
here as $|Q/q_s|$). The constraint radius $R$ was also varied to sample various
curvature values of the ``protein", and thus various total bending angles of
the confined ``DNA", Fig.~\ref{fig:confine}.

\paragraph{A nucleosome model.}
For the nucleosome model, the system described above was modified to mimic the
confinement of DNA around realistic histone core. The DNA fragment size was
increased to 147 bp, and the non-bonded interactions between monomers were
turned on for an additional realism\cite{Fenley2010}. The fragment was confined
around a cylinder of fixed diameter $R = \sim$100\AA{}, and the walls were
placed $\sim$50\AA{} apart (approximate dimensions of the nucleosome
complex\cite{Fenley2010}).
\section{Results and Discussion}
\subsection{The proposed unified framework of polymer bending}
%
%
    We begin with a useful analogy from classical thermodynamics that connects
system's stability to convex properties of its governing potential. For
example, for a system to be stable against a macroscopic fluctuation in energy,
the entropy of the system as a function of energy, $S(E)$, must be concave
(non-convex). Any chord connecting two points on a graph of $S(E)$ must lie
below the curve itself in order to satisfy the second law of thermodynamics
(maximum $S$). Conversely, the inverse function $E(S)$ must be convex. If,
however, $E(S)$ is not convex over some region, the system phase-separates once
this region is reached, with the properties of the two phases corresponding to
the end points of the convex hull of the non-convex region. The actual,
physical average energy of the system follows the convex hull, which makes the
energy manifestly convex. This very general reasoning, with appropriate choice
of the perturbation coordinate and potential, is applicable to phase
transition of single species polymers (Flory-Huggins Theory\cite{Flory1953}), 
as well as to stretching of
polymers\cite{Savin2013} and other materials\cite{Babicheva2013}.
Here we use the analogy to develop a general framework that describes polymer
response to bending, weak and strong, on the same footing.

\begin{figure}
\centering
\includegraphics[width=0.8\linewidth]{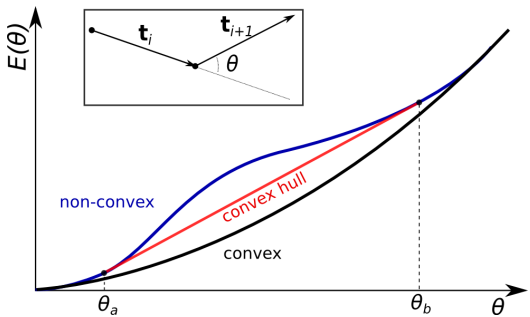}
\caption{Two different forms for a bending energy profile of a homopolymer.
Shown is the (effective) bending energy per site $E(\theta)$. If the
profile
is purely convex down (black curve), the minimal energy conformations of
the
polymer is uniform bending (all sites are identically bent). If the
function
has a non-convex region (blue curve), non-uniform bending is more
energetically
favorable. In this case the total energy of the system follows the convex
hull
of the energy curve (red line).}
\label{fig:conhull}
\end{figure}

    Consider a polymer chain made of $N \gg 1$ inextensible, identical monomer
segments with effective bending deformation energy $E(\theta_i)$ for each
bending site, where $\theta_i$ is the angle between two successive segments
(see inset of Fig.~\ref{fig:conhull}). 
Here we assume that the effective $E(\theta)$ takes into
account all the interactions, short- and long- range, between the
monomers.
For notational simplicity, in what
follows we ignore the difference between $N$ and $N-1$ for large $N$.  
The total energy of the polymer is
$E_{chain}=\sum_i^N E(\theta_i)$, and without loss of generality we assume
no intrinsic bends, \emph{i.e.} $E(0) = 0$. Just like in WLC, we assume
isotropic bending energy, which is a reasonable assumption for DNA fragments
longer than 2 helical repeats or 20 bp\cite{Vologodskii2013}. For the moment,
we further assume no torsional degrees of freedom. In order to induce an
average non-zero bend in the chain, the polymer must be constrained, and the
problem of finding the equilibrium polymer conformation is reduced to
minimizing $E_{chain}$, subject to the specific constraint of the problem.
Here we assume that entropic effects are relatively small at length scales of
interest ($\lesssim L_p$) -- an assumption that we explicitly confirm below by
numerical experiments.

%

	We begin by considering a very special case of a uniformly bent polymer -- constrained to have the same
constant curvature along the entire chain. By construction, such a polymer
consists of identically bent segments with each bending angle $\theta_i$ 
equal to the
average deformation angle, $\bar\theta=N^{-1}\sum_i^N\theta_i$, and its total
energy is $NE(\bar\theta)$. 

	Next, consider a more realistic situation where the polymer bending is
enforced by a much less restrictive constraint: that the sum of the bend angles
between the monomers remains constant, $\alpha=\sum_i^N\theta_i=const$.
Note that
this constraint alone does
not fully define the geometry of
the polymer.  A closed planar loop, with the first and last segments 
linked,  is a relevant
example for which the constraint is satisfied; $\sum_i^N\theta_i = 2 \pi$,
from elementary geometry of polygons, see also the SI.  
Mathematically, the problem of finding the minimum energy conformation
of the polymer is that of energy minimization under the specific
constraint:

\begin{equation}
E_{chain} = N \mathbfcal E(\bar\theta) = \min_{\sum_{i=1}^N\theta_i = N\bar\theta = \alpha} \{ \sum_i^N E(\theta_i) \}
\label{eqn:lagrange1}
\end{equation}
where we make a clear distinction between $\mathbfcal E$, which is 
the average bending energy per bending site {\it in the minimum energy state of 
the polymer}, and $E$ corresponding to the uniform bending. 
Using Lagrange multipliers, Eq.~\ref{eqn:lagrange1} can be reduced to
$\min\{E(\theta_1) +\dots+ E(\theta_N) - \lambda(\theta_1 + \dots + \theta_N -
\alpha) \}$. Differentiating with respect to $\theta_i$ gives a set of
equations $\partial_{\theta_i}E(\theta_i) - \lambda = 0$ (for all $i$) which
leads to a set of equalities $\partial_{\theta_1}E(\theta_1) =
\partial_{\theta_2}E(\theta_2) = \dots = \partial_{\theta_N}E(\theta_N)$. For
a convex functional form of $E(\theta)$, $\partial_{\theta}E(\theta)$
monotonically increases with $\theta$, and therefore the equalities are
satisfied only if $\theta_1 = \theta_2 = \dots = \theta_N$: the polymer 
is always uniformly bent, that is each segment is bent through the same 
angle $\theta = \bar\theta$ and $\mathbfcal E(\bar\theta) = E(\bar\theta)$.
However, for a
non-convex function such as one shown in Fig.~\ref{fig:conhull}, 
there can be more than one value of $\theta$  that 
satisfies $\partial_{\theta_1}E(\theta_1) =
\partial_{\theta_2}E(\theta_2) = \dots = \partial_{\theta_N}E(\theta_N)$: 
$\partial_{\theta_i}E(\theta_a) = \partial_{\theta_i}E(\theta_b)$ 
for some $\theta_a < \theta_b$.  
Of special importance are $\theta_a$ and $\theta_b$
that mark the beginning and the end of the convex hull of $E(\theta)$ 
-- the segment of a straight line tangent
to the non-convex function at two points, such that for any argument 
between these two points
the value of the function at the argument 
is greater than that  of the convex hull line at the same argument, 
Fig.~\ref{fig:conhull}. One can demonstrate, see SI, 
that for bend angles  $\bar \theta$ in the convex hull interval, 
$\theta_a < \bar \theta < \theta_b$, 
a uniformly bent chain is no longer the 
stable minimum energy conformation of the polymer.  
Instead, the stable minimum is achieved when 
the distribution of bend angles
is {\it bi-modal}: each segment 
is bent through one of the
two bending angles $\theta_a$ or $\theta_b$. This general point 
is illustrated in SI for a model 
polymer chain described by a
non-convex bending potential relevant to the case of the DNA.   

	In what follows we derive an explicit expression for 
$\mathbfcal E(\theta)$ for $\theta_a < \bar \theta < \theta_b$. 
In the minimum energy conformation, let $0<p<1$ represent the fraction of all
the bending sites that are in the state $\theta_b$ and $1-p$ the fraction of the
remaining sites in the state $\theta_a$. The total
bending angle in terms of $\theta_a$ and $\theta_b$ is then 
given by $Np \theta_b +
N(1-p)\theta_a = N\bar\theta = \alpha$, and the bending energy per
monomer in the non-convex region is $\mathbfcal E(\bar\theta) = p E(\theta_b) +
(1-p)E(\theta_a)$.
Rewriting $p = (\bar\theta - \theta_a)/(\theta_b-\theta_a)$, we arrive at
\begin{equation}
\mathbfcal E(\bar\theta) = \frac{\bar\theta - \theta_a}{\theta_b-\theta_a}(E(\theta_b) - E(\theta_a)) + E(\theta_a)
\label{eqn:eline}
\end{equation}
Therefore, in the non-convex region, the actual  polymer energy per bending site, $ \mathbfcal  E(\bar\theta)$
corresponding to the stable minimum energy state, 
is a linear
function of the {\it average} deformation $\bar\theta$. Clearly, 
$\mathbfcal E(\bar\theta) < E(\bar\theta)$ within the convex hull interval, 
Fig.~\ref{fig:conhull}. 
 

	To arrive at a general theory that can account for both the
weak and strong bending regimes simultaneously, we use the form of
Eq.~\ref{eqn:eline}
for the strong bending regime, while retaining WLC for the weak bending. In
the
proposed \chlong (\ch) model, the average per segment (e.g. per base-pair)
bending
energy is described by an everywhere differentiable
piece-wise polynomial function: quadratic WLC
(Eq.~\ref{eqn:ewlc}) for $\bar \theta<\theta_a$, and a linear function
 -- convex
hull of $E(\theta)$ -- for $\theta_a<\bar \theta<\theta_b$:
\begin{equation}
\mathbfcal E(\bar \theta) =
\begin{cases}
\frac{1}{2}k_{B}T L_p {\bar \theta}^2 & \text{if } \bar \theta \leq
\theta_a \\
\\
k_{B}T L_p \theta_a (\bar \theta - \frac{1}{2}\theta_a) & \text{if }
\theta_a < \bar \theta < \theta_b
\end{cases}
\label{eqn:echc}
\end{equation}
where $L_p$ is the accepted persistence length, well established for the
weak bending regime; here it dimensionless, expressed in terms of the number of
bending sites (\emph{e.g.} number of base pairs for DNA loops).
In this work we are not interested in
the extreme strong bending regime $\bar \theta>\theta_b$, since for the
DNA this regime would
correspond to loops smaller than 10 bp. Such small loops 
are likely physically impossible due
to steric constraints, and are much smaller than those observed in cyclization
studies\cite{Ha2012,Lionberger2011}. Thus, the only key parameter 
that \ch theory inherits 
from the input effective bending energy, 
$E(\theta)$ in Fig. \ref{fig:model_idea}, 
is the value of $\theta_a$, which enhances robustness of the theory  to inevitable
imperfections\cite{Du2005} of the input bending energy profile. For example, a uniform 
re-scaling $E(\theta) \rightarrow \lambda E(\theta)$ would leave the x-coordinates $\theta_a$ and $\theta_b$ of 
the convex hull double-tangent segment 
unchanged because the derivatives would be re-scaled by the same $\lambda$. Further 
discussion of the robustness of \ch model to its parameters can be found below and in SI.

\subsection{Bending of a circular loop, weak and strong}
\begin{figure}
\centering
\includegraphics[width=0.8\linewidth]{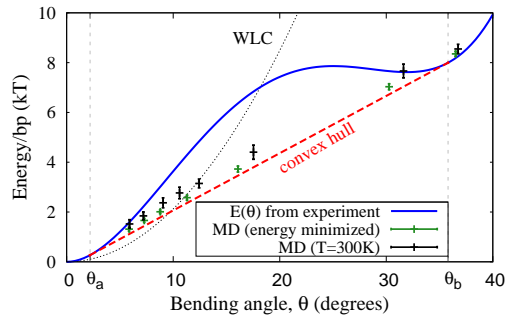}
\caption{DNA effective
bending energy $E(\theta)$ (per bp) extracted from the probability
distribution\cite{Du2005} of DNA bends that naturally occur in protein-DNA
complexes (blue line), and the average energy of unrestrained DNA closed loops
simulated via coarse-grained MD with the same $E(\theta)$ (crosses). Green
symbols: energy minimized (simulated annealing) loops. Black symbols: loops
simulated at T=300K (the corresponding angular probability distribution and example structures are given in the SI). In both cases, the average loop energy as a function of
average bend angle $\bar\theta = \theta$ follows the convex hull of
$E(\theta)$. The small deviation of the T=300K points from the convex hull are
a result of ensemble average sampling and insignificant out-of-plane bending
seen in the simulation. }
%
\label{fig:model_idea}
\end{figure}
%

  While many different types of constraints can be physically realized, one
of the most important ones is the closed loop constraint, which is also
used in DNA cyclization experiments\cite{Baldwin1981,Du2005,Cloutier2005}
critical\cite{Wiggins2005} 
for investigating the strong bending regime. Consider the case of a single
closed loop $\alpha=\sum_i^N\theta_i= 2\pi$.
From Eq.~\ref{eqn:echc}, the total
bending energy of a closed loop of total length $L$ (number of base pairs, corresponding to ``N" in Eq. \ref{eqn:lagrange1} ) is given by
$\mathbfcal E_{loop} = \mathbfcal E(\bar \theta) L$. Since $\bar \theta = \frac{\alpha}{L} = \frac{2 \pi}{L}$, the
bending energy of the loop is:
\begin{equation}
   \mathbfcal E_{loop}(L) =
  \begin{cases}
    2 \pi^2 k_{B}T \frac{L_p}{L} & \text{if } L  > \frac{2 \pi}{\theta_a} \\
    \\
    k_{B}T L_p \theta_a \Big( 2\pi - \frac{1}{2} L \theta_a \Big)  & \text{if }   \frac{2 \pi}{\theta_b}  <  L < \frac{2 \pi}{\theta_a}
  \end{cases}
  \label{eqn:echcloop}
\end{equation}
Note that, where defined, the new function $\mathbfcal E_{loop}$ 
depends on just one new
parameter: $\theta_a$ -- lower boundary of the non-convex domain.
Although we tacitly assumed the loop to be
confined to a 2D plane to simplify the derivations, our
unconstrained coarse-grained
simulations of closed loops at 300K demonstrate, Fig.~\ref{fig:model_idea}, that 
the assumption has little effect on our key conclusions.



\subsection{Application to double-stranded DNA}
    The preceding discussion was not restricted to the case of DNA:
non-uniform, two-phase  
bending, and
the corresponding linear bending regime can be a feature of any polymer. However, since DNA
is arguably the most important polymer, and it exhibits looping in many
different biological systems, we will focus on double-stranded DNA for the rest
of the study. An effective bending energy (per bp) calculated from a
statistical analysis of experimental PDB structures of DNA-protein complexes
\cite{Du2005} is shown in Fig.~\ref{fig:model_idea}. This effective bending
energy function has a non-convex region, and thus a convex hull, the end points
of which are $\theta_a=2.2^{\circ}$ and $\theta_b=35.8^{\circ}$, corresponding
to fragment lengths of $L \sim 160$ and $\sim 10$ bp respectively for DNA
closed loops.
%

    Coarse-grain molecular dynamics simulations at 300K (see Supplementary
Material) demonstrate that polymers with this effective bending energy between
monomers exhibit all of the key features discussed above. For large loop sizes
the bending angles are small (weak bending) -- the system samples the convex
(harmonic) region of the energy function, Fig.~\ref{fig:model_idea}, and the
distribution of bend angles is uni-modal. However, as the loop size decreases,
the average angle per bending site $\bar\theta$ increases, eventually crossing
the $\theta_a$ threshold. Once this happens, the energy of the system per bending site
increases linearly with $\bar\theta$,
and the distribution of bend angles becomes bi-modal, until the system reaches
the upper boundary of the convex hull at $\theta_b$.

\subsection{Comparison with DNA cyclization experiments}
    Most experimental cyclization results are expressed~\cite{Cloutier2005,Du2005}
in terms of the Jacobson-Stockmayer {\it j-factor}, which estimates the
probability that a linear polymer of length $L$ forms a closed loop by joining
its cohesive ends~\cite{Baldwin1981,Jacobson1950}. 
While Monte-Carlo based numerical approaches to compute {\it j-factor} 
exist\cite{Sivak2012}, 
here we use a well-established~\cite{Shimada1984,Allemand2006} analytical 
expression for the {\it j-factor} of an unconstrained closed loop:  
\begin{equation}
	j(L)\simeq\frac{k}{L_p^3}\left(\frac{L_p}{L}\right)^5
\exp\left(-\frac{\mathbfcal E_{loop}}{k_B T} + \frac{L}{4L_p}\right)
  \label{eqn:jfact}
\end{equation}%
where
$\frac{1}{L_p^3}\left(\frac{L_p}{L}\right)^5\exp\left(\frac{L}{4L_p}\right)$
accounts for the entropic contribution, averaged over possible looping
geometries, and $\exp\left(-\frac{\mathbfcal E_{loop}}{k_B T}\right)$ is the energy
penalty of bending the DNA fragment to form the loop. We note that the
$k$ depends, in a complex manner, 
on the loop closing geometry and can be expected to
remain invariant over a relatively short range of loop lengths $L$, within the
same experiment. 

\begin{figure}
  \centering
  \includegraphics[trim=0cm 0cm 0cm 3.0cm,clip=true,width=0.8\linewidth]{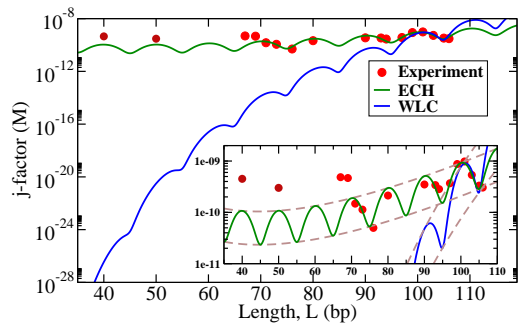}
  \caption{DNA cyclization {\it j-factors} computed using the proposed model
(green line) and WLC (blue line) are compared with recent
experiment~\cite{Ha2012} (red dots, $L>60$ bp).
Experimental values of persistence length, $L_p=150$ bp and $\theta_a=$
2.2$^\circ$ (Fig. \ref{fig:model_idea} ) were used; the value of
$k$ in Eq.~\ref{eqn:jfact} was obtained independently for each model as best
fit against two experimental data points for fragment length L $=101$ and $106$
bp, see Supplementary Material. The envelopes of the {\it j-factor} (brown
dashed lines) for \ch~ and WLC are shown in the inset. Predicted envelope for
\ch {\it j-factor} has a minimum near 45 bp. The experimental data points
$L=50$ and $L=40$ bp were shared by Taekjip Ha (see ref.~\cite{Ha2012}) in
private communication to assess model performance {\it after} the model had
been constructed.}
  \label{fig:jfac}
\end{figure}

	To make a direct connection with cyclization experiments for
non-integer numbers of helical repeats, we modulate the torsionally
independent loop energy from Eq.~\ref{eqn:jfact} with $cos(2\pi L/h)$, 
where we assumed the helical repeat $h=10$ bp per turn. The agreement 
with the cyclization experiment 
is robust with respect to the precise value of the
helical repeat, see SI. 
This form of the modulating factor 
is adopted from Ref.~\cite{Shimada1984} to account
for the periodic variation of the {\it j-factor} due to the torsional component
of the energy~\cite{Shimada1984}. This 
simple way of accounting for non-integer numbers of helical repeats is 
sufficient for the purpose of testing key predictions of \ch vs. WLC, and
does not affect the comparison   
with the over-all (envelope, average) 
behavior\cite{Wiggins2005} of experimental {\it j-factors}, see also Table S1 in SI.   
We use $\mathbfcal E_{loop}(L)$ defined in
Eq.~\ref{eqn:echcloop} for \ch and $\mathbfcal E_{loop}(L)=2\pi^2k_BT\frac{L_p}{L}$ for
all $L$ in the case of WLC. The proposed \ch model and WLC are compared with
the most recent experiment~\cite{Ha2012} in Fig.~\ref{fig:jfac}.

As seen from Fig. \ref{fig:jfac}, \ch~leads to an excellent agreement with the
cyclization experiment, while the $j$-factors predicted by conventional WLC are
off by several orders of magnitude in the strong bending regime (WLC is known
to work well in the weak bending regime where it coincides with \ch by
construction).  The agreement of \ch with the
experiment is robust to the value of its key input parameter $\theta_a$, 
see below and SI.

\paragraph{Cyclization of very short loops.}
	Counter-intuitively, the predicted envelope function for \ch {\it j-factor},
which is essentially Eq.~\ref{eqn:jfact}, has a minimum near 45 bp and begins
to increase for even smaller loops, whereas for WLC {\it j-factor} decreases
sharply for small loops. This completely counter-intuitive behavior of the
cyclization probability for very tight loops predicted by \ch is borne out  
by experiment, Fig. \ref{fig:jfac}; its physical 
origin is explained below. The two experimental points at 
L=50 and L=40, which support the counter-intuitive prediction of
the theory, were not available to us until after the \ch framework was fully
developed and tested against published data\cite{Ha2012} for larger circles. 

	The over-all variation of the {\it j-factor} as a function of the loop length
for both  models is governed by the interplay between the entropic and
the mechanical bending energy costs $\mathbfcal E_{loop}(L)$ of forming the loop. For small
loops, the entropic penalty of forming the loop decreases with the loop size
$L$; however, $\mathbfcal E_{loop}(L) \rightarrow \infty$ for small $L$ within WLC, which
leads to a steep decrease in the over-all cyclization probability. In contrast,
\ch loop energy, Eq. \ref{eqn:echcloop}, approaches a constant for $L
\rightarrow 0$, which explains why the corresponding {\it j-factor} reaches
a minimum and then begins to increase
for small enough $L$, Fig. \ref{fig:jfac}. This very different qualitative
behavior of WLC and \ch {\it j-factors} for small loops can be used as a
discriminating experimental test of the models. The
predicted minimum value of the {\it j-factor} can be used to further
discriminate between models that exhibit the minimum: for example, both 
KWLC\cite{Wiggins2005} and a recent version\cite{Sivak2012} 
of MWLC predict the minimum, but the loop sizes at which the minima occur are
substantially different from the $\sim 45$ base pairs predicted by \ch.

%
Within the proposed \ch framework of polymer bending, the central role is
played by convexity properties of the effective bending energy between
individual monomers. For the DNA, we used the energy profile inferred from
statistical analysis of experimental structures of protein-DNA complexes
(Fig.~\ref{fig:model_idea}) -- the energy has a clear non-convex region,
responsible for the ``softer", linear bending mode of short DNA loops. The
same general considerations will hold for any effective bending energy that has
a distinct non-convex region regardless of its origin\cite{Salari2015},
including a kinkable WLC (kWLC) potential\cite{Vologodskii2013}.
Thus, even though
\ch explains experimental results perceived to be in contradiction
with WLC, there is no fundamental contradiction between the new framework
and the conceptual basis of WLC.

\subsection{Origin of the non-convex bending energy of DNA.} 
\begin{figure}
    \centering
  \includegraphics[width=0.8\linewidth]{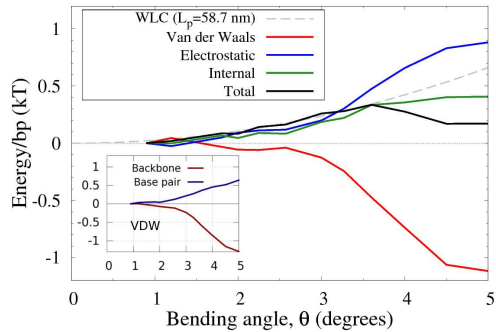}
  \caption{The effective DNA bending energy, per base pair, 
and its physical components as a function of the bending angle $\theta$, 
inferred  from all-atom
MD simulations of DNA circles of variable lengths (50-400 bp). The main
contribution to the non-convexity of the bending energy comes from the
Van der Waals (VDW) interactions. The backbone-backbone part of these
interactions contribute the most to the non-convexity due to a sharp increase
in the attractive energy component for $3^\circ<\theta<4^\circ$, as shown in
the inset. For reference, a WLC fit for the small angle bends ($\approx 1-3.5
^{\circ}$) (grey dashed line) yields the persistence length of 58.2 nm
($\approx 172$ bp), reasonably close to the experimental value of $\approx 50$
nm ($\approx 150$ bp). }
\label{fig:md_energy}
\end{figure}

To investigate, qualitatively, the physical origin of the
non-convexity of the DNA effective bending energy we employed
all-atom Molecular Dynamic
(MD) simulations of uniformly bent DNA circles of a
wide range of sizes, from small to very large, corresponding to almost
unbent DNA, 
see ``Methods". Specifically,
we examined the
average bending energy per base pare.
The total bending energy profile obtained from these simulations, along
with the breakdown into components of different physical origin,
are shown in Fig.~\ref{fig:md_energy}; one can clearly see
a prominent non-convex region, in qualitative agreement with
the experiment, Fig. \ref{fig:model_idea}. The
key parameter $\theta_a \approx 1.5 ^{\circ}$ from the MD simulations,
which is not all that different from the value of $2.2 ^{\circ}$
inferred from the experimental data, Fig. \ref{fig:model_idea}.
Some discrepancy is likely due  to
sequence effects\cite{Lavery2010,WangPettitt2017}, force-field
issues\cite{Cheatham2013,Savelyev2014}, or the fact that the
experiment-based potential in Fig. \ref{fig:model_idea} may itself deviate 
from reality to some extent, as noted in the original
publication\cite{Du2005}. Importantly, the use of MD-derived $\theta_a  =
1.5 ^{\circ}$ in Eqs. \ref{eqn:echcloop} and 
\ref{eqn:jfact}  results, see SI, in virtually the same close agreement with
the cyclization experiment we have seen Fig. \ref{fig:jfac}, which is based
on $\theta_a  = 2.2 ^{\circ}$ derived from experiment. This insensitivity
of the prediction of \ch to the value of its key input
parameter points again to the robustness of the framework.    
The following qualitative conclusions can be made from the MD-based analysis 
of the DNA bending, Fig. \ref{fig:model_idea}. 
For small bending angles, the total energy
is reasonably well approximated by a quadratic function.
However, once
the bending reaches the transition angle $\theta_a$, the VDW
energy decreases at a rate faster than the increase of the other terms
combined, which results in a non-convex region of the total $E(\theta)$,
Fig. \ref{fig:md_energy}. It is this sharp decrease in the VDW contribution
that gives rise to the existence of a non-convex region in the DNA bending
energy. 
Further analysis, (inset in Fig. \ref{fig:md_energy}),
reveals that it is the attractive component of the 
VDW interactions between DNA backbone
atoms (backbone-backbone), rather than base stacking, that is critical to the
counter-intuitive sharp decrease in the total bending energy, see Supplementary Material for further
atomistic details. 
The key role of the backbone-backbone VDW term suggests that it is the overall
structure of DNA, rather than sequence details,
that is responsible for the existence of the convex hull
in the polymer's effective bending profile;
variations in the DNA sequence
may alter the range of bending angles over which the convex hull exists.

	It is worth mentioning that 
local ``bubbles" of broken
WC bonds do not occur in our atomistic MD simulations of DNA circles 
where the uniform bending 
is enforced by constraints on the phosphorous atoms, see ``Methods". 
Yet, these simulations yield a
non-convex profile of the DNA bending energy, 
Fig.  \ref{fig:md_energy}), which, as we have
demonstrated, always 
leads to the existence of linear ``soft" bending regime. Thus, the
simulations 
suggest that local DNA melting 
{\it may not be  necessary}  to explain the high flexibility of
strongly bent DNA and the stark deviation of experimental j-factors from WLC
predictions. We stress that we do not rule out ``bubbles" of broken
WC bonds in actual sharply bent DNA; instead, we predict that
if WC bond breaking were suppressed experimentally, the {\it qualitative } 
picture of sharply bent DNA being much softer would remain, 
with the experimental j-factors still deviating from the WLC in 
a way qualitatively similar\footnote{Quantitative details
can be different if WC bond breaking
is suppressed. Note that 
the effective loop bending energy of \ch theory in Fig. \ref{fig:jfac} 
comes from a statistical analysis of real protein-DNA complexes.
Consequently, the \ch effective energy
with parameters used in that figure, $\theta_a=2.2^{\circ}$ and
$\theta_b=35.8^{\circ}$,  implicitly accounts for broken WC bonds if these
occur in the DNA of the complexes. }  to what is currently observed in experiment, Fig.
\ref{fig:jfac}.  

	An analogy can be made here with the physics behind 
the DNA overstretching
plateau\cite{Smith1996,vanMameren2009}, where the polymer extension occurs at constant force, and the
stretching energy grows linearly with the polymer extension. This peculiar 
regime can be explained\cite{Savin2013} via the same main argument used in the current work -- 
the existence of 
a non-convex region in the polymer deformation energy. 
In the case of DNA
overstretching, experiments
have demonstrated convincingly\cite{Paik2011} 
that WC bond breaking is not required 
for the existence of the characteristic 
plateau on the force-extension diagram.


%
\subsection{Beyond closed loops: a protein-DNA ``complex".}
\begin{figure}
    \centering
  \includegraphics[width=0.8\linewidth]{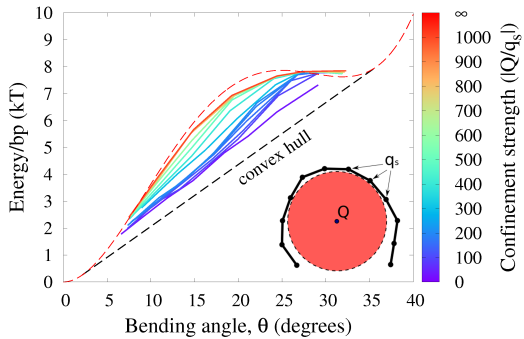}
  \caption{ Polymer bending in a ``protein-DNA complex" model with variable
strength of polymer confinement and curvature, see ``Methods". The red
circle represents the cylindrical charged core of the ``protein" to which
the oppositely charged ``DNA" (black chain) is attracted. Under weak
confinement, the system follows the convex hull of the effective $E(\theta)$,
while approaching $E(\theta)$ (red dashed line) itself for strong confinement.
Shown is the average energy per bead against the average bending angle
$\theta$, at different confinement strengths governed by the ratio $|Q/q_s|$ of
the confining charge $Q$ to the opposite charge $q_s$ of the confined polymer.
The intrinsic bending of the polymer is described by (experimental) $E(\theta)$
from Fig. \ref{fig:model_idea}. }
\label{fig:confine}
\end{figure}

The proposed framework is based on one main assumption: despite constraints,
the polymer chain is still free to explore sufficient conformational space to
search for minimum energy.
So far, we focused on DNA loops because of direct connection to key cyclization
experiments; the single constraint $\sum_i^N \theta_i = \alpha = 2\pi$ is
minimally restrictive. However, other realistic scenarios of DNA bending,
notably in protein-DNA complexes, may involve different types of constraints
that can confine the polymer strongly enough to potentially violate the main
assumption to various degrees. Here we investigate to which extent our main
conclusion -- deformation energy of strongly bent DNA follows the convex hull
of $E(\theta)$ -- may still hold in a model of protein-DNA complex,
Fig.~\ref{fig:confine} and ``Methods".
We vary the total positive charge $Q$ of the cylindrical core to modulate the
electrostatic attraction of the negatively charged polymer (monomer charge
$q_s$) to the core surface of the ``protein", and, hence, the degree of the
polymer confinement.
In the limit of very strong confinement ($|Q/q_s| \rightarrow \infty $), the
polymer is forced to be confined to a circular, uniformly bent path on the
surface of the cylindrical core, and has very few degrees of freedom left to
explore in this regime, solid red line in Fig.~\ref{fig:confine}. The average
bending energy in this case follows the given functional form of $E(\theta)$
(red dashed line in Fig.~\ref{fig:confine}), and \ch clearly does not apply.
As we decrease the confinement strength, the polymer is allowed to
assume non-uniform bending conformations while lowering its total bending
energy. The effective bending energy per monomer begins to approach the convex
hull (solid green and blue lines), making \ch more applicable. In the case of
the weakest confinement (solid purple line), the polymer is still loosely bound
to the core, but is allowed to relax almost completely.
This is the limiting case described by our \ch model: the resulting energy
per monomer follows the convex hull fairly closely

We argue that it is this low confinement regime, where \ch is relevant, that
describes the real nucleosome~\cite{Luger1997} -- arguably the most important
DNA-protein complex. To illustrate, we model a ``variable confinement
nucleosome" by a coarse-grained 147-bp DNA fragment placed next to a cylinder
with relative dimensions of the actual histone core~\cite{Fenley2010}, see
Methods; as the core charge $Q$ is increased, the whole fragment starts to
wrap around the cylinder once the confinement strength is $|Q/q_s|\geq90$.
At this value of the DNA confinement, the energy cost of pulling away a
fragment of $\sim 20$ bp in our model is $\approx 10 k_{B}T$, which is
comparable to $\approx 6 k_{B}T$ estimated from experiment~\cite{Garcia2007}
for the fragment of the same length in the actual nucleosome. Moreover, even
for higher degrees of confinement, up to $|Q/q_s| \simeq 200$ ($\simeq 20
k_{B}T$ to pull away a 20 bp fragment), the corresponding blue lines in
Fig.~\ref{fig:confine} still approximate the convex hull, and so \ch is still
likely applicable, at least qualitatively.

\section{Conclusion}

	It is now well established  that slightly bent DNA behaves like an elastic rod --
the deformation energy is a quadratic (harmonic) function of the deformation.
However, recent experiments demonstrated that strong bending of small DNA
fragments could no longer be described within this classical model. 

	Here we have proposed a novel framework for bending of
polymers, which is based on the consideration of convex properties of the
effective bending energy between successive monomers. Within the framework,
the bending energy is harmonic for small bends, but once the average deformation reaches the convex hull 
of the effective bending energy function, a ``phase transition" to the strongly bent regime
occurs, in which the system's energy is a linear function of
the average bending angle. In this regime, which persists for as long as the average deformation is 
within the convex hull interval, the two states of bending co-exits: some segments are bent weakly, while
others are bent strongly, with the proportion of the latter increasing with the increased average bend 
({\it e.g.} shorter loops). The transition point from the harmonic to the linear
bending regime occurs at the beginning of the convex hull segment -- this point
plays a special role in the new theory.  These
general considerations  are expected to hold for any polymer with an effective bending energy that has
a distinct non-convex region, regardless of its origin\cite{Salari2015,Vologodskii2013}. 

	For generic ``sequence-averaged" double-stranded DNA
considered here, we conclude that the effective bending deformation energy becomes non-convex
for strong bends greater than $\sim2^\circ$/bp, which corresponds to circular
loops shorter than $\sim 160$ bp. The
conclusion about the DNA bending energy being non-convex relies on
an analysis of a large number of
experimental protein-DNA complexes,  and is consistent with the shape of the 
bending energy inferred from atomistic MD simulations. 
The simulations also yield 
a qualitatively similar value for the bend angle that marks the onset of the linear bending regime.
Further, atomistic 
simulations of 
DNA circles reveal that the attractive short-range Lennard-Jones
interactions between the backbone atoms are key for the underlying
non-convexity of the DNA effective bending energy, leading to the linear
bending regime. We use MD simulations only for general reasoning, which is 
robust to details of the simulation
protocol.  

	In this work our focus is the main principle;
future refinements of the \ch theory may be able to account for details not
considered here, such as
sequence dependence of the DNA bending energy, the influence of torsional 
stress and supercoiling, etc. We have also just barely touched upon 
structural consequences of \ch, such as the number and distribution of ``kinks" in tightly bent DNA. An analysis of these features will likely lead to more verifiable predictions of the theory.  Likewise, we
have derived specific mathematical expressions for bending under 
only one type of constraint; other relevant types of constraints need to be
considered in more detail to complete the theory. Based on our analysis, the key
conceptual features of \ch will likely hold.


	 The new theory does not contradict the conceptual basis of the
classical models of DNA bending such as WLC, but also agrees with recent experimental
cyclization data on strongly bent small DNA circles\cite{Ha2012}. A completely
counter-intuitive prediction that cyclization probability reaches a minimum for
very small loops has proved to be consistent with additional experimental data
points, not available to us when we made the prediction.

We believe that the novel general framework 
can be used to analyze, {at least conceptually}, 
many other scenarios of
strong polymer bending, and should help interpret future experimental
observations.

\section{Funding}

This work was supported in part 
by the National Institutes of Health [R01 GM099450] and the National
Science Foundation [MCB-1715207]. 

\section{Acknowledgments}

We thank Taekjip Ha for
sharing with us unpublished experimental data.
We thank Igor Tolokh for many detailed and insightful  comments. 

The authors acknowledge Advanced Research Computing at Virginia Tech for
providing computational resources and technical support that have
contributed to the results reported within this paper.

\section{Author Contributions}
AD, AM and AVO performed the research and wrote the manuscript. AVO designed the research.

\clearpage

{\Large \textbf{SUPPLEMENTARY MATERIAL} } 

\section{Convex vs. non-convex bending energy functions}

Below is an  argument for the difference in bending behavior of polymers
characterized by a convex vs. non-convex bending energy.  As in the main text,
we assume that the polymer deformation energy has only the bending component,
which is isotropic.  In all cases, bending is induced by constraint
$\sum_i^N\theta_i= N\bar\theta = const$, where $\bar\theta$ is the average
bending angle of the polymer chain made of $N$ segments.  

\begin{figure}[bt!]
  \centering
  \includegraphics[width=0.9\linewidth]{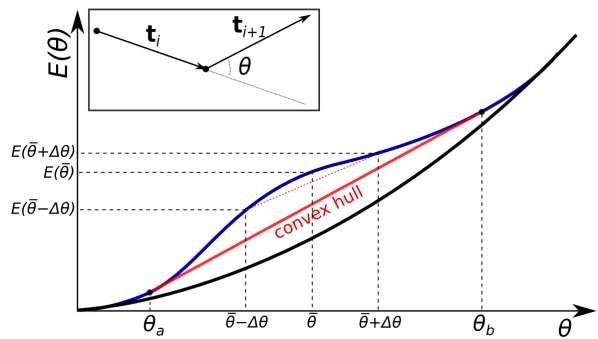}
  \caption{\label{fig:conhull1}Two different forms for a bending energy profile  
of a homopolymer. Shown is the (effective) bending energy per site $E(\theta)$. 
If the profile is purely convex down (black curve), the minimal energy conformations 
of the polymer is uniform bending (all sites are identically bent). 
If the function has
a non-convex region (blue curve), non-uniform bending is more energetically favorable.
In this case the total energy of the system follows the convex hull
of the energy curve (red line).}
\end{figure}

	Without the loss of generality, 
consider the uniformly bent conformation of a polymer with $N=2$, and let's
investigate its stability to perturbation. A perturbation $\Delta\theta$ that
reduces the bend angle at one site means that the other bending site must
increase its bend angle by $\Delta\theta$ in order to satisfy the constraint
$N\bar\theta = const$.  This perturbation changes the total energy $E_{chain}$
by $E(\bar\theta+\Delta\theta) + E(\bar\theta-\Delta\theta) - 2E(\bar\theta)$.
If $E(\theta)$ is a convex function (e.g. the black curve in
Fig.~\ref{fig:conhull1}), the perturbed system will have a higher energy than in
the initial uniformly bent case.  This is because, by definition, a convex
curve always lies below its chords, so that $2E(\bar\theta) <
E(\bar\theta+\Delta\theta) + E(\bar\theta-\Delta\theta)$.  Thus, the
perturbation leads to increase in system energy which implies that the system
was at a stable equilibrium.  Therefore, the minimum energy conformation of a
polymer with a convex effective bending energy is always that of a uniformly
bent chain.

	However, if the function $E(\theta)$ has a non-convex region, e.g. the
blue curve in Fig.~\ref{fig:conhull1}, then for any $\bar\theta$ in that region,
and $\Delta\theta$ that does not take the system outside of it,
$2E(\bar\theta) > E(\bar\theta+\Delta\theta) + E(\bar\theta-\Delta\theta)$ ,
which means it is possible to lower the energy of the polymer further by
non-uniform bending. Namely, one site is now bent through
$\bar\theta-\Delta\theta$, and the other through $\bar\theta + \Delta\theta$,
with the new value of the average chain energy per site, $\frac{1}{2}
E_{chain}$ falling on the midpoint of the line (dashed red line in
Fig.~\ref{fig:conhull1})  connecting the two new bending states on the energy
curve. The longer the cord connecting the two perturbed states at
$\bar\theta-\Delta\theta$ and $\bar\theta+\Delta\theta$, the larger the energy
gain $2E(\bar\theta) - E(\bar\theta+\Delta\theta) - E(\bar\theta-\Delta\theta)$
due to the non-uniform bending, for as long as the cord is completely below the
$E(\theta)$ curve. The largest energy gain, and thus lowest possible
$E_{chain}$, is achieved for the limiting cord that is the convex hull of
$E(\theta)$ -- the line segment tangent to the non-convex function at two
points, such that between these two points the value of the function is greater
than that at any point of the line segment. For this limiting case,
$\bar\theta-\Delta\theta = \theta_a$, $\bar\theta + \Delta\theta = \theta_b$.

\enlargethispage{-65.1pt}

\section{Concave vs. convex closed loops. }

The exterior angle sum theorem is valid for convex polygons,
directly applicable to the sum of the tangent vectors of a convex closed curve.
We do not consider non-convex (concave) closed curves here because these can
not minimize the deformation energy of a closed inextensible elastic loop, at
least as long as the effective bending energy $E(\theta)$ is a monotonic
function of $\theta$, or curvature $\kappa$. (The existence of a non-convex
region in $E(\theta)$ does not imply a non-monotonic $E(\theta)$. ) 

\begin{figure}[h!]
  \centering
    \includegraphics[width=0.9\linewidth]{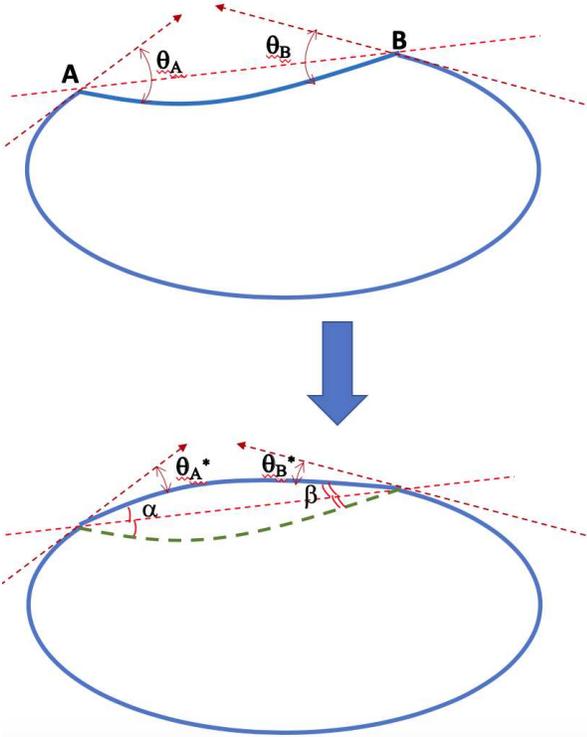}
  \caption{The bending energy of the concave closed curve (top) can be
reduced by reflecting the concave portion of the curve across the convex hull line
($|AB|$, red), to produce a fully convex curve (bottom). 
The procedure reduces the bend angles $\theta_A$ and $\theta_B$ to,
respectively, $\theta_A^* = \theta_A -  2\alpha$
and $\theta_B^* = \theta_B - 2\beta$, while keeping the curvature unchanged
everywhere else along the curve. For a monotonic $E(\theta)$ the net result is a lower bending energy of the loop.   } 
 \label{fig:bangout}
\end{figure}

	The outline of an intuitive proof idea is as follows.  
Any concave curve can be transformed -- ``banged out" -- to eliminate a local
non-convex region while bringing the bending energy down in the process, see
Fig.~\ref{fig:bangout} for an illustration of the process for a shallow
``dimple".  For a deeper ``dimple", the procedure may involve two steps. 
First, the entire convex portion of the curve is ``stretched out" via 
a uniform re-scaling 
of its curvature  $\kappa (s) \rightarrow \lambda \kappa (s), 0 < \lambda
< 1$. The bending energy of the convex portion of the
curve will become lower during this step.
Since the curve is inextensible, and there are no breaks, points A
and B will move further apart as a result, 
thus also reducing the curvature of the
non-convex ``dimple" that spans the $|AB|$ segment, and hence lowering its
bending energy.  After that, the step in Fig.~\ref{fig:bangout} is applied,
making the resulting curve convex, and lowering its bending energy further.  We
do not pursue a more formal proof  based on variational calculus.

\section{Non-convex bending energy function leads to bi-modal distribution 
of bending angles}

\begin{figure}[h!]
  \centering
    \includegraphics[width=0.9\linewidth]{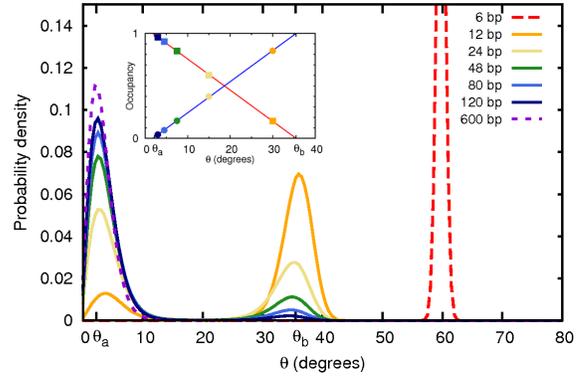} 
  \caption{Angular probability distributions at 300K  
resulting from the non-convex bending potential (Fig. 2 of the main text) used
in  coarse-grained simulations of DNA closed loops of variable size.  As the
loop size (indicated in the top right corner) decreases, the average bending
angle per base pair increases. When the average angle falls into the convex
hull range, the angular distribution becomes bi-modal with peaks at $\theta_a$
and $\theta_b$, corresponding to the weakly and strongly bent states, 
respectively.  The bending through the larger of the two
values, $\theta_b$, can be interpreted as ``kinking".  The fractional
occupancy of both of these states of bending is shown in the inset as a
function of the average bend angle $\bar\theta$. Circles: occupancy of 
the strongly bent state. Squares: occupancy of the weakly bent state.
Out-of-plane motion likely affects angular probability 
distribution of the largest
(600 bp) loop, which may explain the shift, compared to expectation, 
of the position of the 
corresponding distribution peak. } 
 \label{fig:angdist}
\end{figure}

One clear consequence of non-convexity of the bending potential
(Fig.~\ref{fig:conhull1}), is that the corresponding distribution of polymer
bend angles becomes bi-modal once the average bending angle $\bar \theta$ is
within the convex hull region. A weak enough bending is always
uniform, for as long as the average bend angle $\bar \theta$ is below
$\theta_a$. As the average bend angle becomes just slightly larger than
$\theta_a$, most of the segments are still bent weakly through $\theta_a$, but
a small fraction becomes strongly bent through $\theta_b$. As the constraint
forces the system to bend further, the fraction of the strongly bent segments
increases linearly with $\bar \theta$, until, eventually all the segments are
strongly bent through $\theta_b$. Beyond that point the system re-enters the
uniform bending regime again.  Here we confirm this expectation
quantitatively, for a coarse-grained DNA model with one bead per bp.
Specifically, we have analyzed angular distribution of different sized loops,
from 6 to 600 bp long. The closed loops were created and simulated as discussed
with the Main text.  The loops were simulated at $T=300K$ to create 100,000
snapshots. The corresponding  probability distribution of bend angles for  each
loop size is shown in Fig.~\ref{fig:angdist}.

Note that for a large, but finite 
$N \gg 1$, the predictions in Fig.~\ref{fig:angdist} should be 
interpreted as applicable to the entire conformational ensemble  
of bent loops; in particular, the predicted linear dependence of the
fractions of strongly and weakly bent fragments apply to
the ensemble averages. For a given value of $\bar \theta$, between $\theta_a$ 
and $\theta_b$, one will observe a distribution of tightly bent fragments
among the loops, a given loop can have none or more than one;
bent conformations we observed in our MC simulations
were qualitatively consistent with the above picture.
In this work we did not pursue a detailed analysis of ``structural" 
consequences of \ch model, in part because of an uncertainty associated with
the value of $\theta_b$ appropriate for double-stranded DNA. We hope to
revisit this issue in the future.

\begin{figure}[h!]
  \centering
    \includegraphics[width=0.4\linewidth]{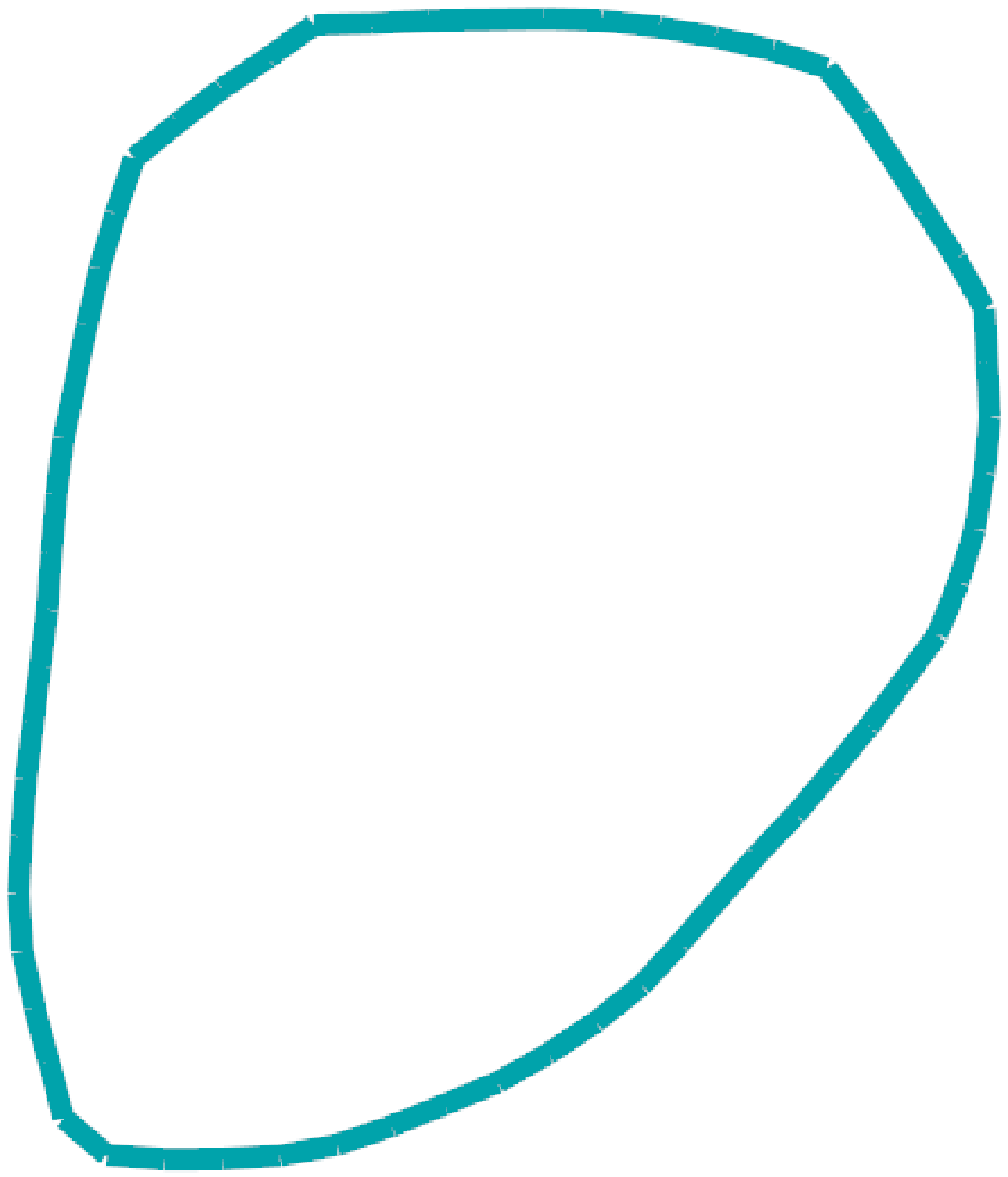}
    \includegraphics[width=0.4\linewidth]{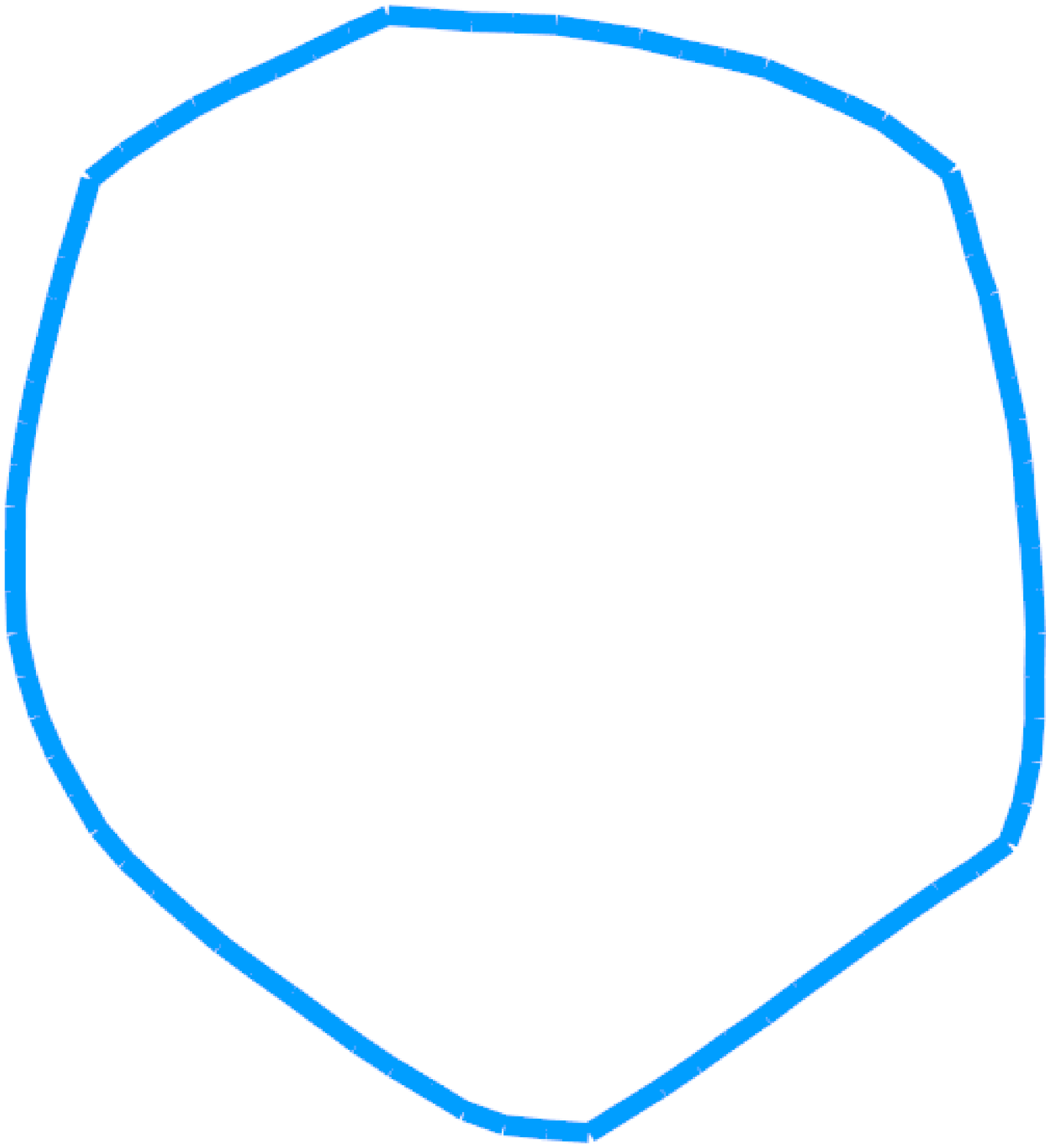}
  \caption{Examples of a 60- base-pair (left) and  80- base-pair (right) loop
from the conformational ensemble shown in Fig. \ref{fig:angdist}. } 
 \label{fig:loopexamples}
\end{figure}

\section{j-factor envelope functions and fitting to experiment}
The difference between \ch and WLC models is encoded in the {\it j-factor} 
via 
\begin{equation}
	j(L)\simeq\frac{k}{L_p^3}\left(\frac{L_p}{L}\right)^5
\exp\left(-\frac{\mathbfcal E_{loop}(L)}{k_B T} + \frac{L}{4L_p}\right)
  \label{eqn:jfact1}
\end{equation}%
where $\mathbfcal E_{loop}$ is the bending energy of forming a loop within each
model, see Main Text. For quantitative comparison with experiment we follow the
standard approach and augment the above formula with an oscillatory term that
accounts for periodic variations in $J(L)$ due to torsional rigidity. 

Specifically, the torsional dependence
of the {\it j-factor} is represented by a cosine function, similar to that of the original
work~\cite{Shimada1984}, to oscillate between the upper and lower envelope functions, 
with a period  of $h=10$ bp, see Fig. 3 in the Main Text. 

\begin{align}
  J(L) & = \frac{1}{2}\Big[J_{top}\Big(1+\cos\Big(\frac{2\pi L}{10}\Big)\Big) \nonumber \\
  & + J_{bot}\Big(1-\cos\Big(\frac{2\pi L}{10}\Big)\Big) \Big]
\label{eqn:cosjfactor}
\end{align}
where the ``top"/``bottom"  envelope curves have the functional form of the
{\it j-factor} in Eq.~\ref{eqn:jfact1},
%
%
\begin{equation}
J_{top/bot}(L) = \frac{k_{top/bot}}{L_p^3}\left(\frac{L_p}{L}\right)^5 e^{\left(L_p\theta_a\left(2\pi - \frac{1}{2}L\theta_a\right) + \frac{L}{4L_p}\right)}
\label{eq:chcenvelope}
\end{equation}

Here we are only  
interested in loops of length $L<160$ bp, since for $L > 160$ bp  \ch = WLC by construction for DNA.  
Thus, the curve representing \ch in Fig. 3 of the main text is Eq. \ref{eqn:cosjfactor}, 
with the corresponding  top/bottom
envelope functions given by Eqs. \ref{eq:chcenvelope}. 

For the case of WLC 
we replace $\mathbfcal E_{loop}(L)$ with the appropriate expression,
so $J_{top/bot}$ used in Eq.~\ref{eqn:cosjfactor} are now given by:

\begin{equation}
J_{top/bot}(L) = \frac{k_{top/bot}}{L_p^3}\left(\frac{L_p}{L}\right)^5 e^{\left(2\pi^2\frac{L_p}{L} + \frac{L}{4L_p}\right)}
\end{equation}

\begin{figure}[htbp]
\centering
   \begin{subfigure}[b]{0.48\textwidth}
   \includegraphics[width=0.95\textwidth]{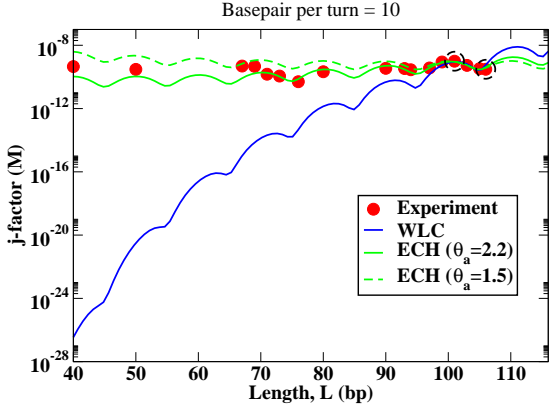}
   \caption{}
   \label{fig:Ng1} 
\end{subfigure}
\begin{subfigure}[b]{0.48\textwidth}
   \includegraphics[width=0.95\textwidth]{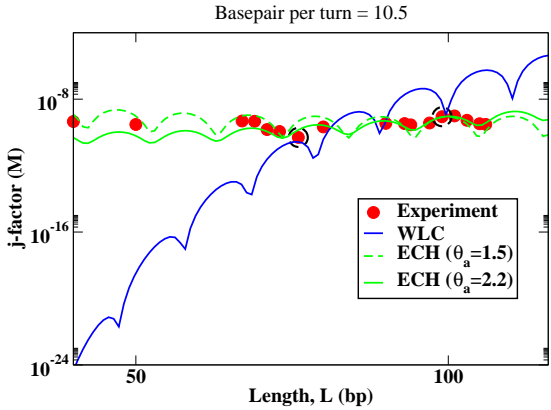}
   \caption{}
   \label{fig:Ng2}
\end{subfigure}
\caption{Robustness of \ch model predictions (green curves ) 
to its input parameters. In (a), \textit{j-factors} are 
estimated using base-pair per helical repeat $h=10$, and 
in (b) $h=10.5$. In each panel, the \ch predictions are based on two
different values of the key input parameter. Solid green lines 
correspond to  $\theta_a = 2.2 ^{\circ}$
inferred from experiment, and dashed green lines correspond to $\theta_a
= 1.5^{\circ}$ from MD simulations. The fitting procedure is the same as
described above, the two experimental 
points used to obtain the asymptotes are indicated by dashed black circles.
For (a) these points are the same as in the main text.  }  
  \label{fig:jfact-robust}
\end{figure}

The values of $k_{top}, k_{bot}$ in the above equations are inferred from
fitting  Eq.~\ref{eqn:cosjfactor} to experimental $J(L)$ points;  since we have
added an oscillatory part to $j(L)$ of Eq.~\ref{eqn:jfact1}, we need two
parameters in the functional form of $J(L)$.  For the fit we have chosen two
experimental data points for loop lengths $L=101$ bp (the {\it top} envelope of
$j(L)$)  and $106$ bp (the {\it bottom} envelope). The reason we chose $J(L)$
that correspond to the largest $L$ available is because the resulting fit
presents the most stringent test for the model in the tight bending regime
where $L$ is small. Note that each curve in 
Fig. 3 of the Main Text, and Figs. \ref{fig:Ng1}, \ref{fig:Ng2}, has its own set of  best fits values of $k_{top}, k_{bot}$, which explains why the WLC and \ch curves do not coincide in the limit of large loops, where \ch approaches WLC. 

\begin{table}[tbh!]
  \caption{\small {\it j-factor} ratios, $J(L_1)/J(L_2)$, predicted using \ch and WLC models compared with experiment\cite{Ha2012}.}
  \label{tab:jfactexp}
    \begin{tabular*}{\linewidth}{@{\extracolsep{\fill}}p{0.08\linewidth}p{0.08\linewidth}p{0.13\linewidth}p{0.13\linewidth}p{0.13\linewidth}@{}}
    \toprule
    $L_1$(bp) & $L_2$(bp) & Experiment & \ch & WLC \\
    \colrule
    40 & 50  & 1.50       & 0.993      & 1.12\e{-6}\\
    71 & 101 & 1.51\e{-1} & 2.01\e{-1} & 3.08\e{-6}\\
    80 & 101 & 2.17\e{-1} & 3.28\e{-1} & 6.10\e{-4}\\
    90 & 101 & 3.56\e{-1} & 5.59\e{-1} & 3.73\e{-2}\\
    \botrule
  \end{tabular*}%
\end{table}

	The \ch envelope functions, Eqs.~\ref{eq:chcenvelope},  have a minimum, which 
can be derived directly from the expression of j-factor (Eq.~\ref{eqn:jfact1}).
The minimum occurs at
\begin{equation}
\centering
L=\frac{5}{\frac{1}{4L_p}+\frac{L_p\theta_a^2}{2}}
\end{equation}
Using the value of $L_p=150$ bp and $\theta_a=2.2^\circ$, the minimum of {\it j-factor} is found at $L\simeq45$ bp.
No such minimum exists in the WLC case in the range of $L$ of interest to us. 

Note that ratios of J-factors at integer values of helical repeats can be predicted directly from 
Eq.~\ref{eqn:jfact1}, which does not contain the oscillatory components. The constant $k$
also cancels, which presents a convenient way to compare \ch theory directly to experiment, 
Table~\ref{tab:jfactexp}.
%

%
%
%
%

\section{Robustness to model details}
\begin{figure}[!h]
  \centering
  \includegraphics[width=0.9\linewidth]{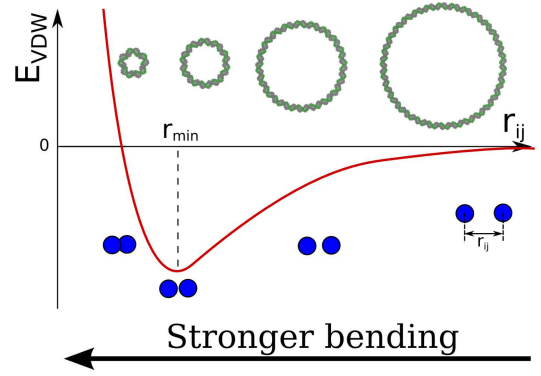}
  \caption{The increase of the bending angle of the DNA duplex causes a
decrease in the average distance between atom pairs that contribute
significantly to the total VDW $E_{VDW}$energy, modeled here as the
Lennard-Jones (LJ) potential; the atoms move deeper into the LJ potential well.
The shape of the well is conducive of a sharp decrease in the total VDW energy
upon small changes in the atom-atom distance $r_{ij}$ caused by the DNA
bending.  Once the average distance passes the LJ well minimum, the VDW energy
begins to increase again. }
  \label{fig:vdwbend}
\end{figure}

The key parameter of the \ch theory is the transition point,
$\theta_a$ from the quadratic to the linear bending regime. As seen from Fig.
\ref{fig:jfact-robust}, a good agreement with the experimental {\it j-factors}
is achieved over a relatively broad  range of $\theta_a$ values. Specifically,
both $\theta_a=2.2^{\circ}$ inferred from experiment, and
$\theta_a=1.5^{\circ}$ obtained from atomistic MD simulations (see Main Text)
yield nearly the same agreement with experiment.  In both cases, the
counter-intuitive prediction of increased cyclization probability for very
short loops hold.  


Predictions of \ch framework are also robust to the precise
value of the helical repeat parameter $h$ used in Eqs. \ref{eqn:cosjfactor} and
\ref{eqn:jfact1} to account for the torsional stress created by loops with
non-integer number of helical repeats. The use of $h=10.5$, more appropriate
for DNA in solution\cite{Wang1979}, vs. $h=10$ corresponding to classical
B-form, has little affect on the agreement of the predicted {\it j-factors}
with the experiment.

\section{Origin of the non-convex bending at the atomic level} 

 To illustrate, at the atomic level, the origin of the non-convex behavior of
the backbone-backbone VDW energy, consider pairs of oxygen (O) atoms in the
backbone. We choose these atoms because they have one of the lowest VDW energy
minima out of all atom pairs in the DNA, and have most atom pairs within the
effective short range of the interaction. The O-O VDW energy as  a function of
pairwise distance $r_{ij}$ has a potential well at $r_{min} \sim 3.3\AA{}$,
Fig.~\ref{fig:vdwbend}.  Further  analysis of pairwise interatomic distances
reveals that as the double helix bends, the geometry of the backbone deforms in
a way that more oxygen atoms pairs fall into this well ($\sim3.1-3.5$\AA{}):
beyond $\theta_a$ the accumulation of the attractive contributions begins to
sharply lower the total interaction energy compared to the native, unbent state
where all the oxygen atoms are significantly further apart ($>$4.0\AA{}) than
the well minimum. As the helix bends further, these O-O pairs eventually pass
the LJ minimum, and begin to climb onto the repulsive wall, which increases the
total VDW interaction energy, as seen in the corresponding figure of the main
text.  

\enlargethispage{-65.1pt}

\bibliographystyle{apsrev4-1}
\bibliography{bending_jcp}







\end{document}